\documentclass[aps,prl,superscriptaddress,showpacs,twocolumn]{revtex4}
\usepackage{amsfonts}
\usepackage{amsmath}
\usepackage{amssymb}
\usepackage{graphicx}
\begin{document}
\title{Spin liquid phase of
the $S=\frac{1}{2}$ $J_1-J_2$ Heisenberg model on the triangular lattice}
\author{Zhenyue Zhu}
\affiliation{Department of Physics and Astronomy, University of California,
Irvine, California 92697, USA}
\author{Steven R. White}
\affiliation{Department of Physics and Astronomy, University of California,
Irvine, California 92697, USA}

\begin{abstract}
We study the $S=1/2$ Heisenberg model on the triangular lattice with nearest- and next-nearest-neighbor interactions 
$J_1$ and $J_2$ with the density matrix renormalization group, on
long open cylinders with widths up to nine lattice spacings. 
In an intermediate $J_2$ region $0.06 \lesssim J_2/J_1 \lesssim 0.17$, 
we find evidence for a spin liquid (SL) state with short range
spin-spin, bond-bond, and chiral correlation lengths, bordered by a classical $120^\circ$
N\'eel ordered state at small $J_2$ and by a two sub-lattice collinear magnetically ordered state at larger $J_2$.
Focusing on $J_2/J_1 = 0.1$, we find a number of signatures of a gapped SL phase.

\end{abstract}

\date{\today}
 \maketitle
\medskip

In Anderson's paper introducing the resonating valence bond (RVB) state\cite{tr1},
the prototypical example of a spin liquid (SL)\cite{SL},
the ground state of the triangular lattice nearest neighbor Heisenberg model was argued to be a likely candidate.
Later, a variety of analytical and numerical studies\cite{tr2,tr3,sw1,ed} demonstrated that
this system has three sublattice $120^\circ$ long range antiferromagnetic order.
More recent numerical studies\cite{gfmc,se1,dm1} have confirmed this result, and more accurately determined
the magnetization, with $M\sim0.2$. 

It is natural to include small second-neighbor $J_2$ terms to the Hamiltonian,
in addition to the nearest-neighbor terms with coupling $J_1$, to see if this
additional frustration induces a spin liquid state.
The corresponding classical phase diagram
has a single phase transition point at $J_2=1/8$ (setting $J_1=1$ here and below) 
between the $120^\circ$ 
phase and a large number of degenerate four sublattice magnetically ordered states.
This degeneracy is broken by quantum fluctuations within spin wave theory, selecting a two sublattice collinearly 
ordered state through the order by disorder mechanism.\cite{4c,2c} 

One might expect an intermediate phase to appear near the classical critical point at $J_2=1/8$.
The limited number of studies on this question, which have usually relied on approximations with uncertain reliability,
have given conflicting results, particularly on the
nature of a possible disordered phase and the location of the phase boundaries.\cite{sb,vmc1,vmc2,ccm}
Here, we try to resolve the nature of this possible intermediate state using density matrix renormalization
group (DMRG) methods.\cite{dmrg} 
We {\it do find} a spin liquid intermediate phase,  gapped with fairly large singlet and triplet gaps,
which is bordered 
by the expected magnetic phases, the $\sqrt{3}\times\sqrt{3}$ ordered state ($120^\circ$ classical
Neel order pattern) at  
$J_2 < 0.05\sim0.07$ and a two sub-lattice collinear ordered state at $J_2\gtrsim0.17$. 
The SL state away from the phase boundaries at $J_2=0.1$ has very short range magnetic, bond, and 
chiral correlations. We also observe a dimerization pattern of bond strengths on odd cylinders
and obtain two different topological sectors on even cylinders.  This behavior is in a number of ways
similar to that observed in the $Z_2$ spin liquid state of the kagome Heisenberg model.\cite{kag, kag2}
A possible $Z_2$ SL state on the triangular lattice was treated analytically in the early 1990's.\cite{tsl1,tsl2}
However, in contrast to the kagome, it has a strong tendency towards spatial 
anisotropy in the bond strengths. 

\begin{figure*}
\includegraphics*[width=17cm, angle=0]{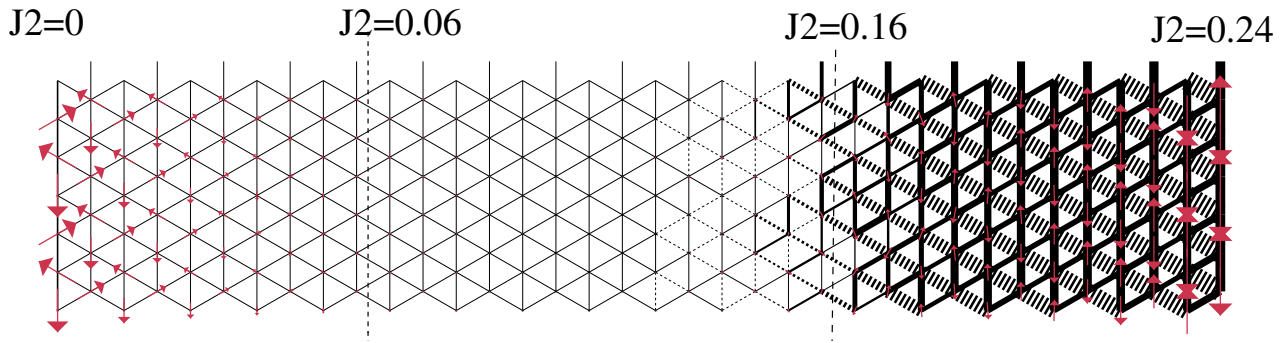}
\caption{(Color online) For a YC6 cylinder, we vary $J_2$ with position, from
$J_2=0$ on the left edge to $J_2=0.24$ on the right edge. We
also apply a pinning magnetic field along both the x and z directions on
the left edge to favor the classical $120^\circ$ order. Two approximate
phase transition lines are shown. 
The size of the arrow represents local measurement of $\langle S \rangle
=\sqrt{\langle S_x\rangle^2+\langle S_z\rangle^2}$ with the
direction of the angle given by $\tan^{-1}[\langle S_z\rangle/\langle S_x\rangle]$, and the widths of lines 
proportional to $|\langle S_i\cdot S_j\rangle+0.18|$. The solid lines
along the bonds mean the bond measurement is negative, i.e., a stronger than average bond,
while dashed lines indicate bonds that are weaker than average.
} \label{phase}
\end{figure*}

We study the Hamiltonian
\begin{equation}
H=J_1\sum_{\langle i,j\rangle}S_i\cdot S_j+J_2\sum_{\langle\langle
i,j\rangle\rangle}S_i\cdot S_j
\end{equation}
where $\langle i,j\rangle$ and $\langle\langle i,j\rangle\rangle$ run over nearest- and next-nearest-neighbor
pairs of sites.  We set $J_1=1$
and consider only $J_2>0$.  
We study open-ended cylinders, with the axis along the x direction.
If one of the three bond directions lies along the x(y) direction, we call it an XC(YC)
cylinder. An XCn cylinder has $n$ sites along the zigzag y-direction, while a YCn cylinder
has a circumference of $n$ vertical bonds.

The triangular lattice, with six $J_1$ and six $J_2$ bonds,
has more connecting bonds than other lattices recently studied with DMRG. This both increases the
number of Hamiltonian terms and increases the
entanglement, which is to first order governed by the area law. 
For example, a vertical line
through the YCn cylinder cuts 2n near-neighbor bonds; thus, one would expect a greater entanglement entropy
in this system than in a square, honeycomb, or kagome lattice with the same width.
This means we have to keep more states $m$ for the same accuracy, while the greater number of 
Hamiltonian terms increases the computational and memory cost for a given $m$.
The widest cylinders that we can calculate accurately  
are YC9 and XC10, keeping up to M=6400 states, which produces
a truncation error that is always less than $10^{-5}$. 

First, we present one calculation which shows all three phases 
along a single cylinder. In Fig. \ref{phase}, we vary $J_2$
spatially from 0 (left edge) to 0.24 (right edge) on a YC6 cylinder,
where we label the possible phase transition points in this model.
At $J_2\lesssim 0.06$, we see the $\sqrt{3}\times\sqrt{3}$ magnetically ordered state, with a diminishing
order parameter as one nears the transition.
For large $J_2$ values, we see a two sub-lattice collinear ordered phase consistently across various cylinders,
which resembles the N\'eel order on a tilted square lattice, consistent with spin wave theory.\cite{2c}

For $0.06\lesssim J_2 \lesssim 0.16$ on this YC6 cylinder, there is a region with very small 
magnetic moments, and with a nearly uniform
nearest-neighbor bond strength pattern.
Below we will study in detail the point
$J_2=0.1$, near the center of the intermediate phase.  We find that all of our results are
consistent with this phase being a gapped SL. 

We now focus on $J_2=0.1$, in the center of the nonmagnetic phase.
To understand the results it is essential to distinguish the different possible 
topological sectors for a finite cylinder with open ends. 
(We consider an even number of sites.)
Infinitely long cylinders are either even or odd, based on the
number of sites in a one-dimensional (1D) unit cell.  For example, a YCn cylinder is even if n is even. 
Call this type of parity C.
In addition, another parity arises based on a near-neighbor dimer picture.  
Given any dimer covering, if we cut the cylinder with a vertical 
line not intersecting any sites, the number of dimers cut gives another parity.
Call this parity D; we also refer to it as the even or odd (topological) sector.
For a finite cylinder, assuming perfect dimer coverings, the D parity is determined 
by how the left and right ends are terminated, and moving a site from the left end to the right (or vice versa) 
switches the topological sector.
In a C-odd cylinder, the two D-parity sectors are related by a translation of one 1D unit cell, so the bulk 
properties are identical.
In a C-even cylinder, the two D-parity sectors are significantly different, but the bulk properties 
become identical as the cylinder width increases in a $Z_2$ SL.
For finite width, a ground state of the higher-energy sector may be able to fall into the lower-energy sector,
through the creation of a spinon at each end of the system.
The C parity is a rigorous concept associated with the Lieb-Schultz-Mattis theorem.
It is not obvious that the D parity is a useful concept for every spin liquid, but for both the kagome
and the triangular SL found here, the classification appears to work perfectly.

\begin{figure}
\includegraphics[width=8.5cm, angle=0]{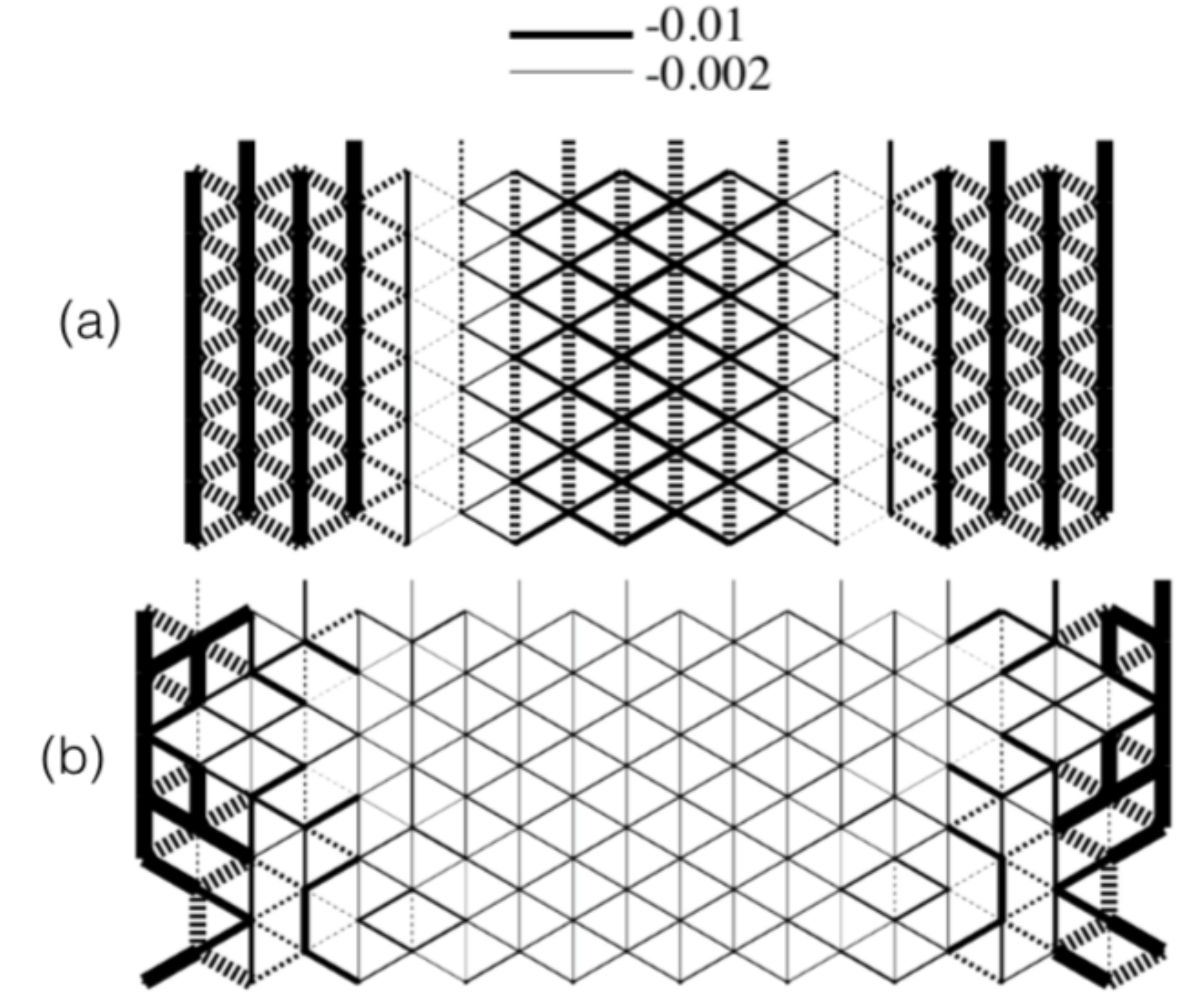}
\caption{(a) The higher-energy even and (b) the lower-energy odd sector ground
states for a YC6 cylinder with $J_2=0.1$, where we subtract $-0.18$ from
all the bonds.  The odd and even sector systems differ primarily 
by the removal of a single site at each edge; in addition, we needed to make
the higher-energy system shorter to avoid falling into the low-energy sector
through the creation of two end spinons.  
In the plot the bond thickness is
restricted to a maximum; otherwise, many edge bonds would be much thicker.  
(c) Central portion of the ground state on the XC6 cylinder. 
The solid (dashed) bonds have strength $\langle S_i\cdot S_j\rangle = -0.287 / -0.157$.
(d) A similar central region for a YC5 cylinder.
The solid (dashed) bonds have strength $\langle S_i\cdot S_j\rangle = -0.158 / -0.126$.
} \label{yc6}
\end{figure}

In Fig. \ref{yc6} we show results for the ground states for both sectors for the (C-even) YC6 cylinder.
Here we see that the lower-energy sector
has a very uniform bond strength pattern (bottom panel), whereas the higher-energy sector
is much less uniform. This behavior is seen in all the C-even cylinders, in both 
this triangular system and in the kagome Heisenberg system,
thought to be a $Z_2$ spin liquid.\cite{kag}  

For a $Z_2$ spin liquid, these two sectors in a C-even cylinder should become degenerate in the
two dimensional (2D) limit, with the energy separation depending exponentially on the width of the
cylinder. Here, for YC6, extrapolating in the truncation error and in the cylinder length,
we find an energy per site for the lower-energy odd sector of $E_0=-0.52096(1)$.  For a long
enough cylinder, the even sector produces end spinons and falls into the odd sector.
The end spinons cost a finite energy, of order of the triplet spin gap, but being
in the wrong sector in the bulk costs an energy proportional to the length of the system.
Thus, short system even sector ground states are stable. Longer systems, during the course
of a DMRG simulation, may stay in the even sector ground state for a number of sweeps, but then
as we increase the number of states kept $m$, they may suddenly fall into the lower-energy
sector by producing two end spinons. (We can also prepare the initial DMRG state to make it start
off in the two spinon sector, in which case there is no sudden fall.)
For example, for a YC6 cylinder with length $L_x=30$, we have observed a sudden drop near $m \sim 3000$,
but this depends on a variety of details of the DMRG simulations.
Thus, estimating the higher-energy ground state energy cannot be done as accurately
as the low-energy sector. 
(The DMRG calculations also converge faster and with
smaller truncation errors for the lower-energy sector.)
Using shorter cylinders, for YC6 we find an even
sector energy of $E_1=-0.5152(2)$, higher than the odd sector by about 0.0058(2)
per site, or about $1.1\%$.
The magnetic correlations, the bond-bond correlations, and the chiral correlations
for the YC6 low-energy sector are all very short ranged, with
correlation lengths roughly one to two lattice spacings.\cite{sup}

Similar behavior is seen for the C-even YC4 and YC8 cylinders. However, whereas for YC6 the
bond strengths in the three bond directions were almost identical, for YC4 they
are highly anisotropic.
For YC4, the ground state is in the even sector,
while the odd sector energy is 
higher by about $3\%$. In the even sector, the diagonal bond strength ($-0.045$) is 
almost ten times weaker than the vertical bond strength ($-0.442$). 
In the odd sector, the opposite happens: The diagonal bond ($-0.23$) 
is three times larger than vertical bonds ($-0.08$).
It appears that this spin liquid state is highly susceptible to bond anisotropy,
and the small circumference of the YC4 cylinder elicits very large anisotropic responses.

On the YC8 cylinder, the ground state is in the odd sector with an energy 
$0.6\%$ lower than the even sector. 
The odd sector has uniform bond strengths in the cylinder center, but as YC4
it has a significant bond anisotropy, with a vertical bond strength of -0.225 and a horizontal bond strength of
-0.159. (This strong tendency towards anisotropy on such a large lattice is completely absent
in the kagome system.) The higher-energy even sector has nonuniform bond strengths,  appearing as if 
there are strings connecting two ends of the cylinder.\cite{sup}

Comparing YC4, YC6, and YC8, we see that the energy difference between the two
sectors falls steadily with increasing width. 
For a gapped $Z_2$ SL, 
the energy splitting should decay exponentially with
increasing the cylinder width.  
Our results are consistent with this exponential decay,
with a decay length of about 1.7 lattice spacings (not shown).
The YC cylinders can have significant bond anisotropy, although for YC6 it is very small.  
Comparing YC4 and YC8, the strength of the anisotropy falls rapidly with width, while
for YC6 it is anomalously small.

On the C-even XC cylinders, such as XC4 and XC8, anisotropy is also
observed. (XC6 is an odd cylinder, so we will discuss that below.) 
With the XC cylinder geometry, finite size effects make the horizontal
bonds weaker than the two diagonal bonds.  The anisotropy is less
pronounced on XC8 than on XC4.

We have tried to measure the
topological entanglement entropy to more directly measure the topological order for the SL
state.  However, because of the strong anisotropy, the entanglement entropy
for various cylinders cannot be linearly extrapolated versus the cylinder
width for our current range of widths.

For C-odd cylinders, the dimer picture predicts two degenerate ground states which differ
only by a horizontal translation, thus obeying the Lieb-Schultz-Mattis theorem.
These two states are always visible in our results through bond strength distortions, as they
are for the kagome SL.
These distortions decrease in intensity with cylinder width, as expected.
Figures \ref{yc6}(c) and \ref{yc6}(d) shows the dimerization patterns on the XC6 and YC5 cylinders. 
Similar dimerized patterns are also observed on all other C-odd XC and YC
cylinders.

To quantify the bond distortion, we define the dimerization order parameter $D$ as the 
difference between the strong and weak bonds along the two diagonal directions, for both YC and XC
cylinders. We find that, for all
C-odd cylinders, $D$ is almost constant in the cylinder center, indicating
long range dimerization order,
and decreases for wider cylinders.
In contrast, on C-even cylinders, $D$ decays
exponentially away from the left and right edges.\cite{sup}
The behavior
is quite similar to that of the kagome SL and provides
additional evidence in support that the state is a spin liquid.

\begin{figure}
\includegraphics[width=8cm, angle=0]{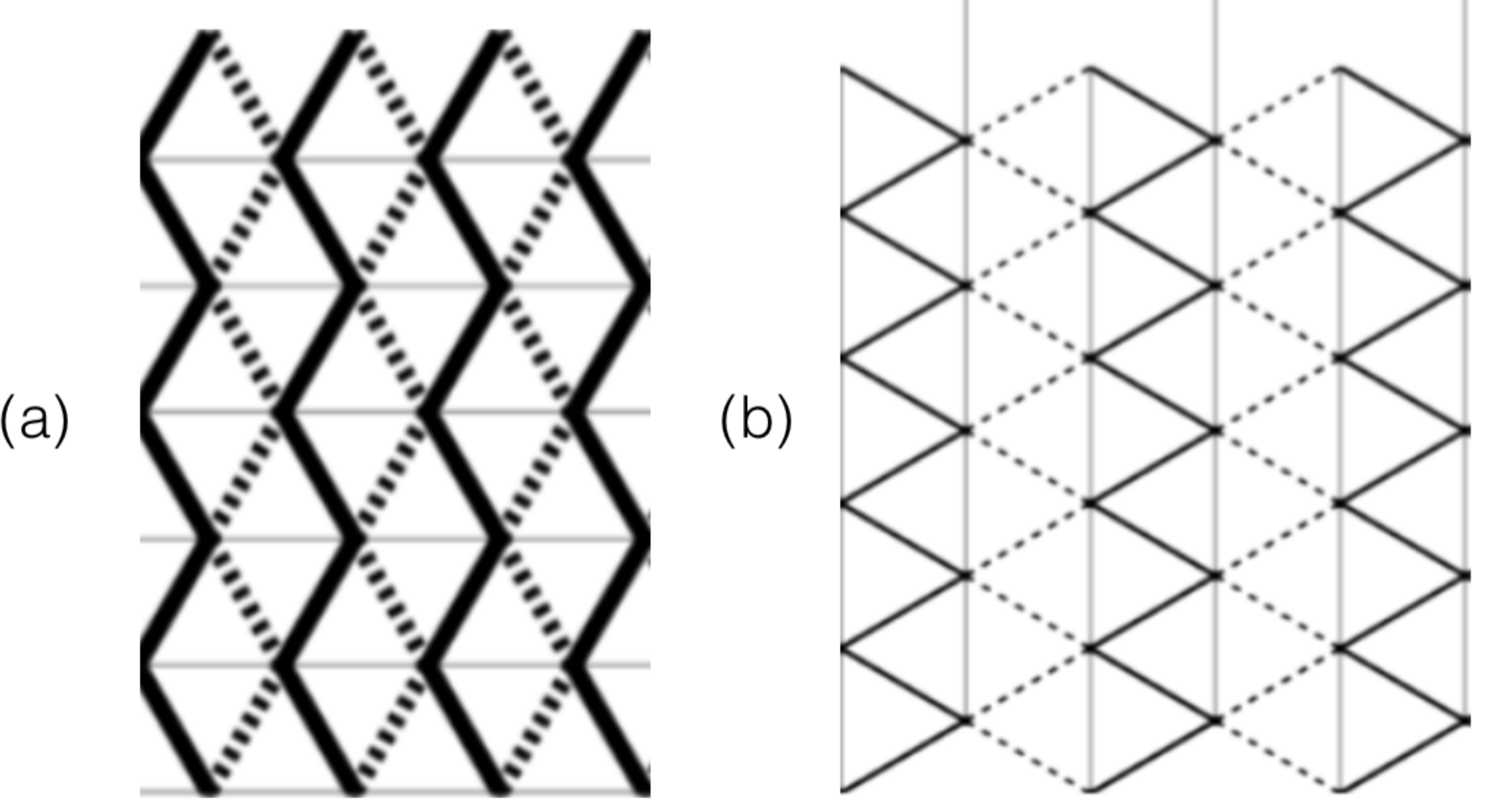}
\caption{
(Color online) The spin triplet gap for various long cylinder geometries at $J_2=0.1$ with $L_x=20$.
The inset shows the linear extrapolation of spin triplet gap for YC6 cylinders vs
$1/L_x^2$. The triplet gap is roughly $\Delta_T=0.357$ for an infinitely long YC6 cylinder.
}
\label{gap}
\end{figure}

We display results for triplet spin gap in Fig.  \ref{gap} for $J_2=0.1$.
The gaps are typically two to three times as large as that of the 
kagome system ($\Delta_T \sim0.14$, \cite{kag2}).  The gaps show a relatively minor
finite size behavior, compared to their magnitude.  Each of these gaps in the main part of the
figure is for $L_x=20$; one should extrapolate these to $L_x \to \infty$, and the inset shows this 
extrapolation for YC6.  The gap is proportional to $1/L_x^2$, as expected for a 
simple massive particle (e.g., a particle in a 1D box). The correction to the $L_x=20$ results is
small and we expect that the main figure would only be slightly changed if it used extrapolated results.
Note that for wider cylinders, we need to constrain the spin excitation to the cylinder center, since
otherwise low-energy edge excitations might hide the bulk gap (again, as one must do for the kagome).
A conservative estimate for the bulk 2D triplet gap would be 0.3(1) for $J_2=0.1$, 
and it is hard to imagine it being zero.

One can look at the bond and particularly the spin patterns for the lowest-energy triplet excitations.  On 
even cylinders, the spin excitation resembles a single particle, which we might interpret as
two tightly bound spinons.
However, on odd cylinders, the spin excitation typically looks more as two separate spinons, and seems
more complicated than on even cylinders, with more $L_x$ dependence of the gap.\cite{sup} 
We have only calculated the spin singlet gap for the YC6 and YC8 cylinders. In the kagome system, 
the singlet gap is small, about $0.05$. Here, it is much larger:
$\Delta_s=0.30$ for YC6 and $\Delta_s=0.26$ for YC8. 
Overall, our results strongly support a fully gapped SL state, instead of the gapless SL state in Ref.
\cite{vmc2}.

The finite gaps, short correlation lengths, and topological sector behaviors are all qualitatively
similar to the kagome system and strongly indicate a gapped spin liquid.  However, the
directional anisotropy of the bonds apparent in most cylinders is unlike the kagome, and
raises the question of whether it persists in the 2D limit---which would make it a ``nematic spin liquid"\cite{nsl}.
To try to understand the finite size effects associated with bond anisotropy,  
we have studied systems where we strengthen all the near-neighbor exchange couplings $J$ along
one particular direction and measure the response in the spin-spin correlation pattern.
For the normally isotropic YC6 cylinder, increasing the $J$'s
along one diagonal direction by $5\%$ increases the corresponding bonds
by roughly $30\%$ and decreases the other diagonal bonds by roughly $30\%$---a rather large
response.  For the XC8 cylinder, which is normally quite anisotropic, if we strengthen the $J$'s on the weaker 
(horizontal) bonds by about $3\%$, the weaker bonds increase by about $50\%$ and the system becomes 
approximately isotropic.
These results indicate a large susceptibility associated with a tendency towards 
nematicity. This tendency is a key property of this system, independent of whether
the system actually breaks rotational symmetry in the thermodynamic limit.
Reference \cite{nsl} theoretically studied the phase transition between a $Z_2$ SL state
and different valence bond solid (VBS) orders on a triangular lattice.  They found that
the transition from a columnar or resonating plaquette VBS order can be either
first order or there could be two transitions with an intermediate phase.  The
intermediate phase would host a nematic $Z_2$ spin liquid that breaks 
$2\pi/3$ lattice rotation symmetry.  
Is the triangular system in such a nematic spin liquid state? The anisotropy generally decreases with
system width in the system sizes we can study, but the behavior is irregular, and the effects still large
on the largest widths. Answering this question will require future studies on larger systems.

In summary, we conclude that there is a gapped spin liquid state in the triangular lattice 
Heisenberg model with next-nearest-neighbor exchange $J_2=0.1$. This phase is 
bordered by a three sublattice $120^\circ$ N\'eel ordered state
at $J_2 < 0.05 \sim 0.07$ and a two sublattice magnetic collinear ordered state at $J_2\gtrsim0.17$.\cite{sup2} 
This phase has fairly large gaps, very short correlation lengths, and topological behavior very similar to that
seen in the kagome Heisenberg spin liquid, although we have not been able to measure whether there is
a topological entanglement entropy.
Unlike the kagome, the system has a strong tendency towards nematicity, and whether rotational symmetry
is broken in two dimensions, making it a nematic spin liquid, remains to be determined.

\begin{acknowledgments}
We would like to thank David Huse, Subir Sachdev, Sasha Chernyshev, Yuan-Ming Lu, Yuan Wan, Yi
Zhang, Kevin Slagle, and Miles Stoudenmire for many helpful
discussions.  This work was supported by NSF Grant No. 
DMR-1161348 and by the Simons Foundation through the many electron collaboration.
\end{acknowledgments}

Recently, we became aware of two papers working on the same model with DMRG \cite{tdm1,tdm2}, 
where the spin liquid state is also found \cite{tdm2}.The classification of $Z_2$ spin liquid states is 
subsequently proposed in Refs. \cite{tz1,tz2}.

\end{document}